\documentclass [12pt] {article}
\usepackage{graphicx}
\usepackage{epsfig}
\begin {document}
\begin{center}

{\Large {\bf PRODUCTION OF SECONDARIES IN HIGH ENERGY d+Au
COLLISIONS}} \\
\vspace{.5cm}
C. Merino, C. Pajares, and Yu. M. Shabelski$^{*}$ \\
\vspace{.5cm}
Departamento de F\'\i sica de Part\'\i culas, Facultade de F\'\i sica, \\ 
and Instituto Galego de Altas Enerx\'\i as (IGAE), \\ 
Universidade de Santiago de Compostela, Galicia, Spain \\
E-mail: merino@fpaxp1.usc.es, pajares@fpaxp1.usc.es

\vspace{.2cm}
 
$^{*}$ Permanent address: Petersburg Nuclear Physics Institute, \\
Gatchina, St.Petersburg 188350 Russia \\
E-mail: shabelsk@thd.pnpi.spb.ru
\vskip 0.5 truecm

A b s t r a c t
\end{center}

In the framework of Quark-Gluon String Model we calculate the
inclusive spectra of secondaries produced in d+Au collisions
at intermediate (CERN SPS) and at much higher (RHIC) energies.
The results of numerical calculations at intermediate energies
are in reasonable agreement with the data. At RHIC energies
numerically large inelastic screening corrections (percolation effects) 
should be accounted for in calculations. We extract these effects from
the existing RHIC experimental data on minimum bias and central d+Au
collisions. The predictions for p+Au interactions at LHC energy are
also given.   

\vskip 1cm

PACS. 25.75.Dw Particle and resonance production

\newpage

\section{\bf Introduction}

The Quark--Gluon String Model (QGSM) and the Dual Parton Model (DPM)
are based on the Dual Topological Unitarization (DTU) and describe
quantitatively many features of high energy production processes,
including the inclusive spectra of different secondary hadrons, their
multiplicities and multiplicity distributions, etc., both in
hadron--nucleon and hadron--nucleus collisions at fixed target energies,
as well as the main features of secondary production at collider energies. 
The model parameters were fixed [1-7] by comparison of the calculations 
with experimental data. 

The inclusive densities of different secondaries produced in $pp$ 
collisions at $\sqrt{s} = 200$~GeV in midrapidity region were reasonably 
described in \cite{AMPS}. In the present paper we calculate in the QGSM 
the inclusive spectra of secondaries produced in d+Au collisions both at 
intermediate (CERN SPS, $\sqrt{s_{NN}} = 19.4$ GeV) and much higher 
(RHIC, $\sqrt{s_{NN}} = 200$ GeV) energies. We also present some predictions 
for LHC energies. 

In the QGSM high energy interactions are considered as proceeding via the 
exchange of one or several Pomerons, and all elastic and inelastic processes
result from cutting through or between Pomerons \cite{AGK}. Inclusive
spectra of hadrons are related to the corresponding fragmentation
functions of quarks and diquarks, which are constructed using the
Reggeon counting rules \cite{Kai}.

In the case of interaction with a nuclear target the Multiple Scattering
Theory (Gribov-Glauber Theory) is used, what allows to consider the
interaction with the nuclear target as the superposition of interactions
with different numbers of target nucleons.

The radius of a deuteron is rather large in the strong interaction scale,
so we can assume that the proton and the neutron in the deuteron interact 
independently with heavy nuclei. However, sometimes only one nucleon 
interacts, the second one being a spectator. The average number 
$\langle N_d \rangle$ of the deuteron nucleons interacting with a heavy 
nucleus A in a minimum bias collision is determined by the cross sections 
of secondary production in NA and dA collisions \cite{BBC}
\begin{equation}
\langle N_d \rangle = \frac{A \sigma^{prod}_{NA}}{\sigma^{prod}_{dA}} \,
\end{equation}
and the inclusive spectrum $dn/dy \vert_{dA}$ of any secondary particle
produced in a d+A collision is equal to the product of  $N_d$ times the 
spectrum in a N+A collision:
\begin{equation}
 dn/dy \vert_{dA} = \langle N_d \rangle \cdot dn/dy \vert_{NA} \; .
\end{equation}
In the case of central collisions $\langle N_d \rangle$ can be
calculated with the help of Glauber Theory \cite{CGS}, while the values
of $dn/dy \vert_{NA}$ should be those calculated for central N+A
collisions \cite{Sh2}.

%In the case of heavy ion collisions the Multiple Scattering Theory
%also allows to consider this interaction as the superposition of
%separate nucleon--nucleon interactions. However, in this case there
%is no possibility to sum up all the diagrams in a rather simple form.
%The first simple classes of diagrams can be accounted as the simple
%expressions \cite{CM,Alkh}. The situation with more complicate
%diagrams is not so clear \cite{Andr}--\cite{BoKa}.

%Really, the probability of double interactions in high energy inelastic 
%$pd$ collisions is smaller than 15-20\% \cite{BLRW}. With this accuracy

\section{\bf Inclusive spectra of secondary hadrons in the \newline
Quark-Gluon String Model}

For the quantitative predictions one needs a model for multiparticle
production and we have used the QGSM in the numerical calculations
presented below.

We consider a nucleon as a system of three gluon strings connected to
three valence quarks and joint at the point called ``string junction'' 
(SJ) [14-16], as it is shown in Fig.~1. Sea quarks are produced by
radiation (possibly, via some non-perturbative mechanism) inside the 
gluon strings.

\begin{figure}[htb]
\centering
\vskip -2.cm
\includegraphics[width=.45\hsize]{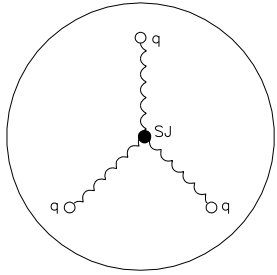}
\vskip -1.cm
\caption{\footnotesize
Composite structure of a baryon in string models. Quarks are shown by open 
points.}
\end{figure}

As mentioned above, high energy hadron--nucleon and hadron--nucleus
interactions are considered in the QGSM as proceeding via the
exchange of one or several Pomerons. Each Pomeron corresponds to a
cylinder diagram (see Fig.~2a) that, when cutted, produces two showers 
of secondaries as it is shown in Fig.~2b. The inclusive spectrum of 
secondaries is determined by the convolution of diquark, valence quark, 
and sea quark distributions, $u(x,n)$, in the incident particles with the 
fragmentation functions, $G(z)$, of quarks and diquarks into secondary 
hadrons. The diquark and quark distribution functions depend on the 
number $n$ of cut Pomerons in the considered diagram.

\begin{figure}[htb]
\centering
\includegraphics[width=.5\hsize]{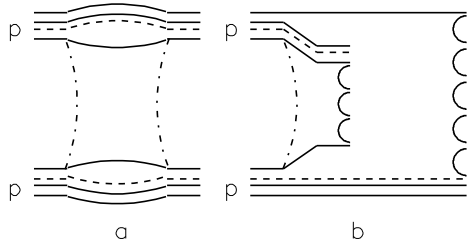}
\vskip -.5cm 
\caption{\footnotesize Cylinder diagram (cylinder is shown by
dash-dotted curves) corresponding to the one--Pomeron exchange
contribution to elastic $pp$ scattering (a), and the corresponding
cut diagram which represents its contribution to the inelastic $pp$ cross 
section (b). Quarks are shown by solid curves and string junctions by 
dashed lines.}
\end{figure}

In the case of a nucleon target the inclusive spectrum of a
secondary hadron $h$ has the form \cite{KTM}:

\begin{equation}
\frac{x_E}{\sigma_{inel}} \frac{d\sigma}{dx_F}
= \frac{1}{\sigma_{inel}} \frac{d\sigma}{dy}
=\sum_{n=1}^{\infty}w_{n}\phi_{n}^{h}(x)\ \ ,
\end{equation}
where the functions $\phi_{n}^{h}(x)$ determine the contribution of
diagrams with $n$ cut Pomerons and $w_{n}$ is the probability of this
process \cite{TM}. Here we neglect the diffraction dissociation contributions 
of diffraction which are comparatively small in most of the processes 
considered below, since they are important mainly for secondary production 
in the large $x_F$ region that is not significant in the present calculations.

For $pp$ collisions

\begin{equation}
\phi_{pp}^{h}(x) = f_{qq}^{h}(x_{+},n) \cdot f_{q}^{h}(x_{-},n) +
f_{q}^{h}(x_{+},n) \cdot f_{qq}^{h}(x_{-},n) +
2(n-1)f_{s}^{h}(x_{+},n) \cdot f_{s}^{h}(x_{-},n)\ \  ,
\end{equation}

\begin{equation}
x_{\pm} = \frac{1}{2}[\sqrt{4m_{T}^{2}/s+x^{2}}\pm{x}]\ \ ,
\end{equation}
where $f_{qq}$, $f_{q}$, and $f_{s}$ correspond to the contributions
of diquarks, valence quarks, and sea quarks, respectively.

These contributions are determined by the convolution of the diquark and 
quark distributions with the fragmentation functions, e.g.,

\begin{equation}
f_{q}^{h}(x_{+},n) = \int_{x_{+}}^{1}
u_{q}(x_{1},n)G_{q}^{h}(x_{+}/x_{1}) dx_{1}\ \ .
\end{equation}
The diquark and quark distributions, as well as the fragmentation
functions, are determined by Regge intercepts \cite{Kai}.

In the case of nuclear targets we should consider the possibility
of one or several Pomeron cuts in each of the $\nu$ blobs of
hadron-nucleon inelastic interactions, as well as cuts between
Pomerons. For example, for a $pA$ collision one of the cut Pomerons 
links a diquark and a valence quark of the projectile proton with a 
valence quark and diquark of one target nucleon, while other Pomerons
link the sea quark-antiquark pairs of the projectile proton with 
diquarks and valence quarks of other target nucleons and also with
sea quark-antiquark pairs of nucleons in the target.

In particular, and as one example, one of the diagrams contributing to for 
the inelastic interaction with two target nucleons is shown in Fig.~3. In 
the blob of the proton-nucleon 1 inelastic interaction one Pomeron is cut, 
and in the blob of the proton-nucleon 2 interaction two Pomerons are cut. 
It is essential to take into account all digrams with every possible 
Pomeron configuration and permutation. The process shown in Fig.~3 
satisfies the condition [18-21] that the absorptive parts of the 
hadron-nucleus amplitude are determined by the combination of the absorptive 
parts of the hadron-nucleon amplitudes.

\begin{figure}[htb]
\centering
\includegraphics[width=.5\hsize]{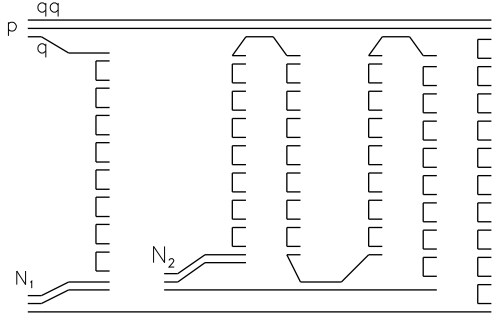}
\caption{\footnotesize One of the diagrams for the inelastic interaction 
of an incident proton with two target nucleons $N_1$ and $N_2$ in a $pA$
collision.}
\end{figure}

Let us consider the case of inelastic interactions with $\nu$ target 
nucleons, $n$ cut Pomerons in $hA$ collisions ($n\geq \nu$), and $n_i$ 
cut Pomerons connecting with the $i$-th target nucleon 
($1 \leq n_i \leq n-\nu+1$), i.e. for the diagram in  Fig.~3 $\nu = 2$,
$n = 3$, $n_1 = 1$, $n_2 = 2$. We define the relative weight of the 
contribution with $n_i$ cut Pomerons in every $hN$ blob as $w^{hN}_{n_i}$. 
For the inclusive spectrum of the secondary hadron $h$ produced in a $p A$
collision we obtain \cite{KTMS}

%\newpage

\begin{eqnarray}
\frac{dn}{dy} =
\frac{x_E}{\sigma^{prod}_{pA}} \frac{d \sigma}{dx_F} & = &
\sum^A_{\nu=1} V^{(\nu)}_{pA} \left\{ \sum^{\infty}_{n=\nu}
\sum^{n-\nu+1}_{n_1 = 1} \cdot \cdot \cdot
\sum^{n-\nu+1}_{n_{\nu}=1} \prod^{\nu}_{l=1} w^{pN}_{n_l}
\right. \times \\ \nonumber & \times &
[f^h_{qq}(x_+,n)f^h_q(x_-,n_l) +
f^h_q(x_+,n)f^h_{qq}(x_-,n_l) + \\ \nonumber & + &
\sum^{2n-2}_{m=1} f^h_s(x_+,n)f^h_{qq,q,s}(x_-,n_m)] \left.
\right\} \;,
\end{eqnarray}
where $V^{(\nu)}_{pA}$ is the probability of pure inelastic
(nondiffractive) interactions with $\nu$ target nucleons, and we
should account for all possible Pomeron permutations and for the
difference in quark content of the protons and neutrons in the
target.

In particular, the contribution of the diagram in Fig.~3 to the
inclusive spectrum is

\begin{eqnarray}
\frac{x_E}{\sigma^{prod}_{pA}} \frac{d \sigma}{dx_F} & = & 2
V^{(2)}_{pA} w^{pN_1}_1w^{pN_2}_2 \left\{
f^h_{qq}(x_+,3)f^h_q(x_-,1)\right. + \\ \nonumber & + &
f^h_q(x_+,3)f^h_{qq}(x_-,1) + f^h_s(x_+,3) [f^h_{qq}(x_-,2) +
f^h_q(x_-,2) + \\ \nonumber & + & 2f^h_s(x_-,2)] \left. \right\}
\;.
\end{eqnarray}

The diquark and quark distributions and the fragmentation functions here 
are the same as in the case of a nucleon target.

\section{\bf Inclusive spectra in p+Au and d+Au collisions at
CERN SPS energy}

The QGSM gives a reasonable description \cite{KTMS,Sh4} of the inclusive 
spectra of different secondaries produced on nuclear targets at energies 
$\sqrt{s_{NN}}$ = 14-30 GeV. 

In Fig.~4 we compare our results obtained by using Eqs.~(2), (7) for the
minimum bias p+Au and for 0-43\% central d+Au collisions with the 
experimental data in the rapidity distributions of negatively charged 
secondaries at $\sqrt{s_{NN}} = 19.4$ GeV \cite{NA35} in laboratory system. 
The calculated value of the average number of interacting beam nucleons in 
these central collisions is $\langle N_d \rangle =$ 1.93. For minimum bias 
interactions we use $\langle N_d \rangle =$ 1.61.

\begin{figure}[h]
\centering
\includegraphics[width=.48\hsize]{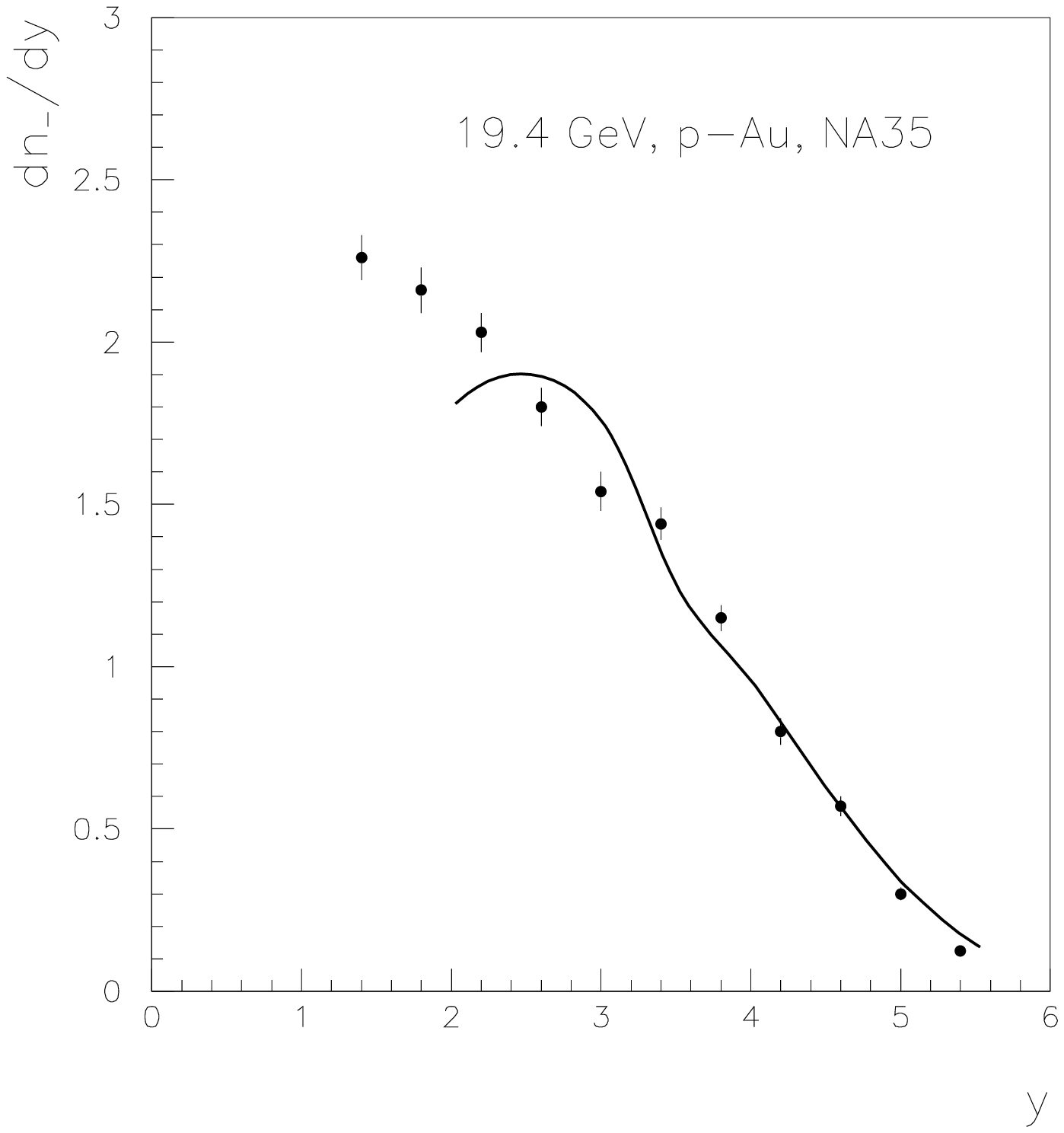}
\includegraphics[width=.48\hsize]{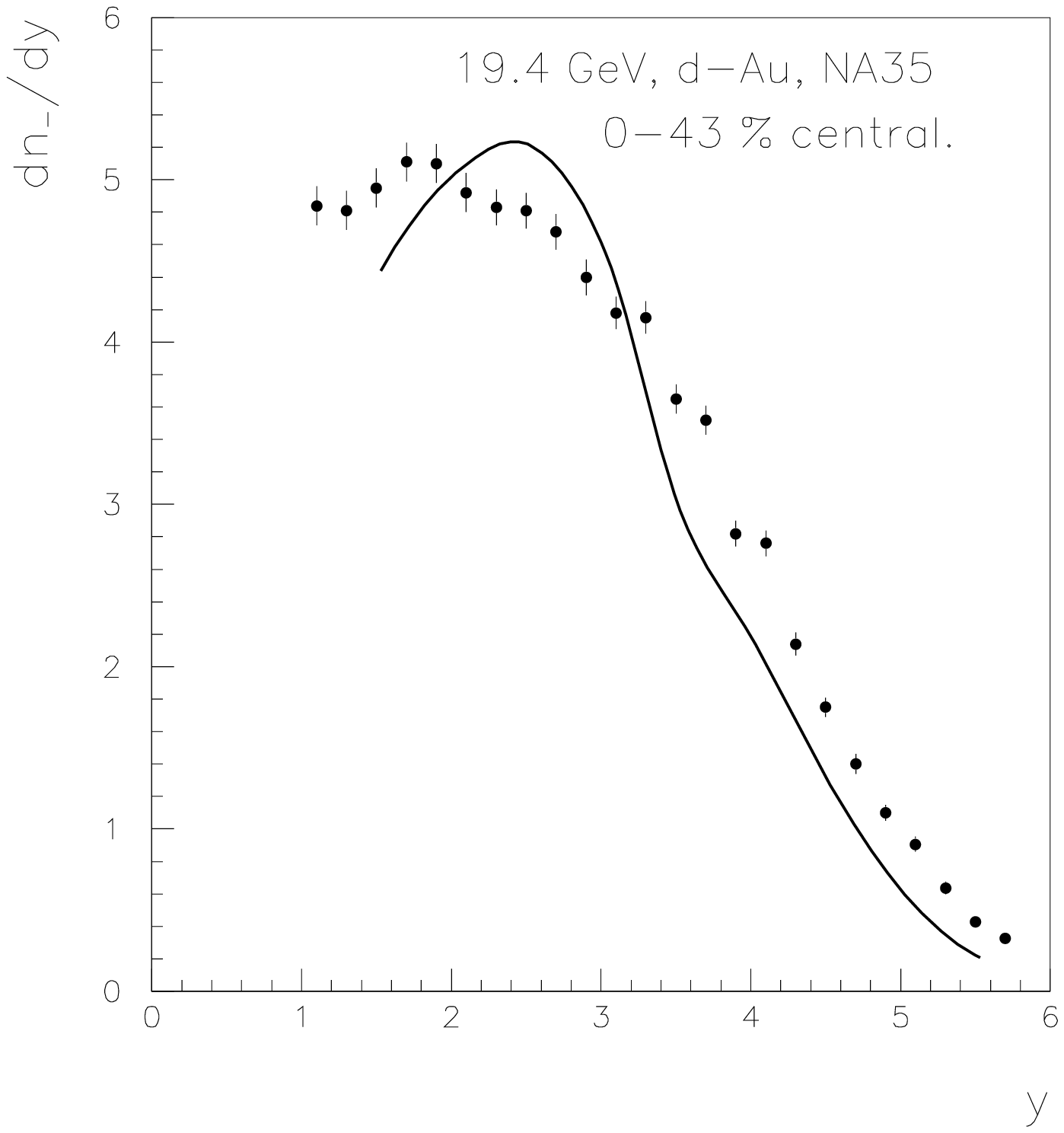}
\vspace*{-0.5cm}
\caption{\footnotesize
Rapidity distributions of negatively charged secondaries at 
$\sqrt{s_{NN}} = 19.4$ GeV \cite{NA35}, measured in minimum bias p+Au  
(left panel) and in 0-43\% central d+Au collisions (right panel) together
with their description by the QGSM. }
\end{figure}

In the central region $y_{lab} \sim 3$, as well as in the beam fragmentation 
region the agreement for both p+Au and d+Au cases is reasonable. The 
differences between the calculated curves and the data are not larger 
than 10\%. In the target fragmentation region,
 (at lab. rapidities 
$y_{lab} \leq 1.5$) our curves underestimate the data, what can be 
connected with Fermi-motion and/or intranuclear cascade contributions.
By comparing these results with those in \cite{KTM,KTMS,Sh,Sh4} one
can see that the A-dependence of the inclusive spectra is reasonably
represented by the QGSM, and that our assumption of independent interaction
of the proton and the neutron in deuteron with the target in d+A collisions 
agrees with the experimental data. 

The NA35 Collaboration has also presented the rapidity distributions of net 
protons and net $\Lambda$-hyperons measured in minimum bias p+Au, and similar 
distributions of net protons in 0-43\% central d+Au collisions. Following 
\cite{ACKS}, for the calculation of these distributions in the QGSM 
we consider three different sources of the net baryon charge. The first one 
is the fragmentation of the diquark, that gives rise to a leading baryon 
carrying the initial SJ (Fig.~5a). A second possibility is to produce a 
leading meson in the first break-up of the string and a baryon in the 
subsequent break-up \cite{Kai,22r} (Fig.~5b). In these two cases the 
baryon number transfer is only possible for short distances in rapidity. 
In the third case shown in Fig.~5c, both initial valence quarks recombine 
with sea antiquarks into mesons $M$, and a secondary baryon is formed by 
the SJ together with three sea quarks.

\begin{figure}[htb]
\centering
\includegraphics[width=.55\hsize]{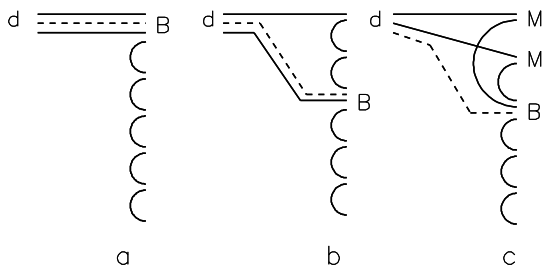}
\vskip -0.1cm
\caption{\footnotesize
QGSM diagrams describing secondary baryon $B$ production by diquark $d$: 
initial SJ together with two valence quarks and one sea quark (a), initial SJ 
together with one valence quark and two sea quarks (b), and initial SJ 
together with three sea quarks (c).}
\end{figure}

The corresponding fragmentation functions for the secondary baryon $B$ 
production can be written as follows (see \cite{ACKS} for more details):
\begin{eqnarray}
G^B_{qq}(z) &=& a_N v_{qq} \cdot z^{2.5} \;,
\\
G^B_{qs}(z) &=& a_N v_{qs} \cdot z^2 (1-z) \;,
\\
G^B_{ss}(z) &=& a_N \varepsilon v_{ss} \cdot z^{1 - \alpha_{SJ}} (1-z)^2
\end{eqnarray}
for the processes shown in Figs.~5a, 5b, and 5c, respectively, and where $a_N$ 
is the normalization parameter, and $v_{qq}$, $v_{qs}$, $v_{ss}$ are the 
relative probabilities for different baryons production that can be found by
simple quark combinatorics \cite{AnSh,CS}. The fraction $z$ of the incident 
baryon energy carried by the secondary baryon decreases from Fig.~5a to 
Fig.~5c, whereas the mean rapidity gap between the incident and secondary 
baryons increases. The contribution of the graph in Fig.~5c contains a 
coefficient $\varepsilon$ which determines the small probability of such 
baryon number transfer. 

The values of the parameters $\alpha_{SJ}$ and $\varepsilon$ which correctly 
describe [28--33] all the data concerning baryon number transfer at high 
energies were presented in \cite{SJ1} :
\begin{equation}
\alpha_{SJ}\, =\, 0.9\;\; {\rm and} \quad \varepsilon\, =\, 0.024\,.
\end{equation} 

In Fig.~6 we compare the results of our calculations of net baryon 
production in p+Au and central d+Au collisions with experimental data
\cite{NA35}. The normalization and general trends are reproduced quite
reasonably. The contribution of SJ diffusion turns out to be more important 
for secondary $\Lambda$ - $\bar{\Lambda}$-hyperon production.

\begin{figure}[h]
\centering
\includegraphics[width=.48\hsize]{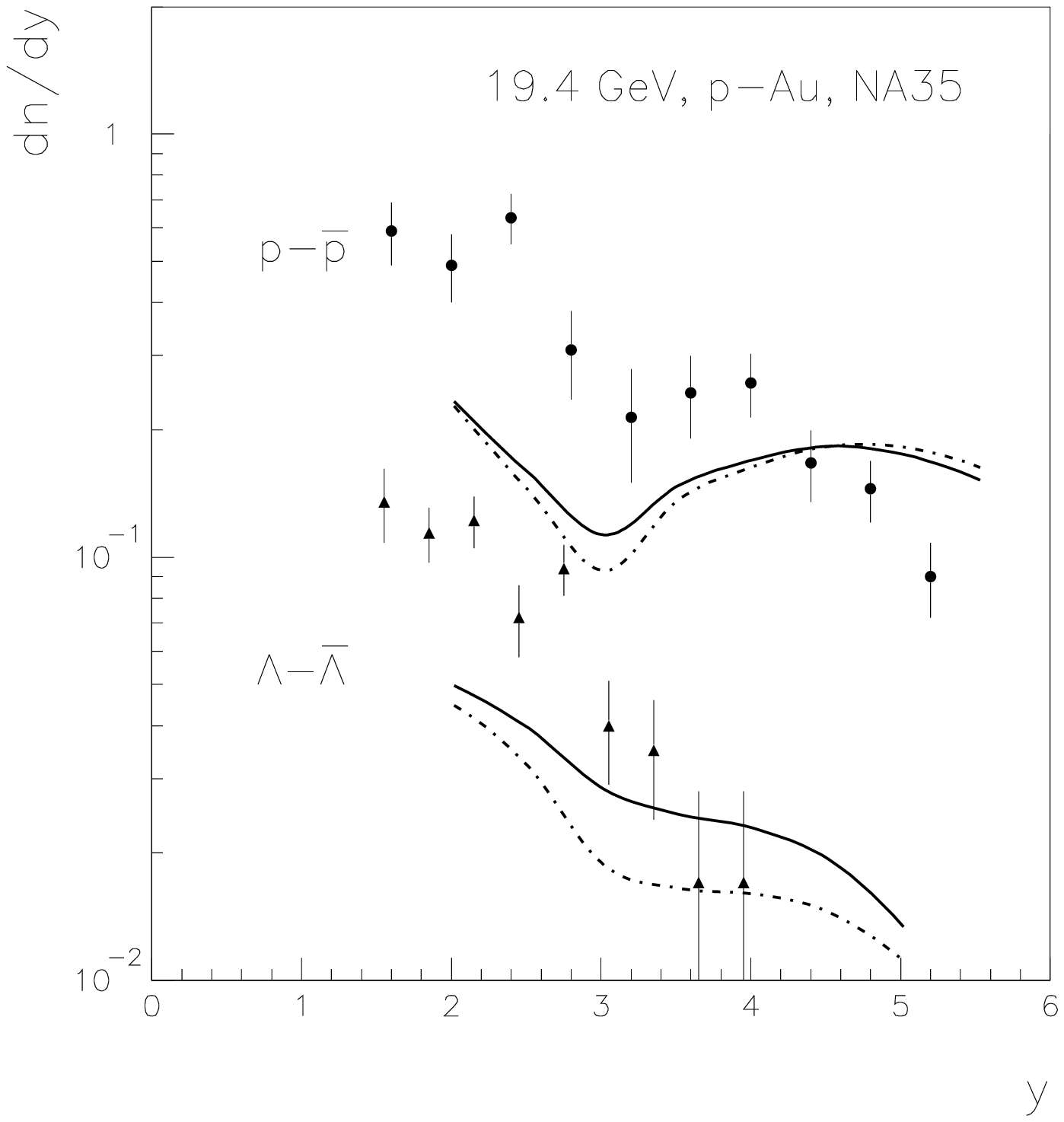}
\includegraphics[width=.48\hsize]{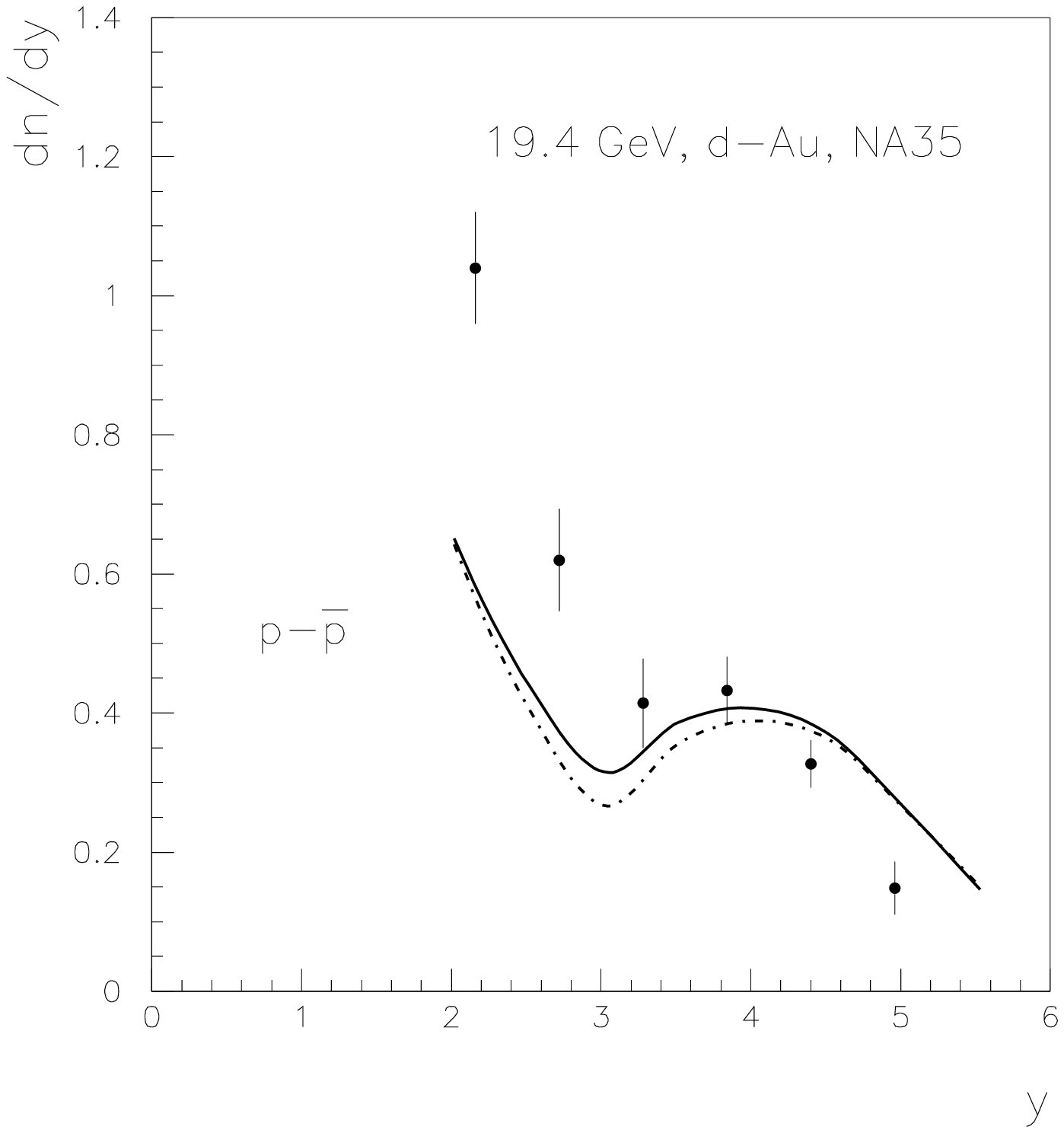}
\vspace*{-0.5cm}
\caption{\footnotesize
Lab. rapidity distributions of net protons and net $\Lambda$-hyperons, 
measured in minimum bias p+Au (left panel) and similar distributions 
of net protons in 0-43\% central d+Au collisions (right panel) at 
$\sqrt{s_{NN}} = 19.4$ GeV \cite{NA35}. The QGSM calculations with
and without SJ contributions are shown by solid curves and dashed 
curves, respectively.}
\end{figure}

\section{\bf Inclusive spectra at RHIC energies}

The c.m. pseudorapidity $\eta^*$ spectra of all charged secondaries in p+p 
collisions at $\sqrt{s_{NN}}$ = 200~GeV were measured by PHOBOS Collaboration 
\cite{PHOB} and these data are in agreement with the results by the UA5 
Collaboration \cite{UA5}. The PHOBOS data are presented in Fig.~7 together 
with the QGSM calculation shown by solid curve. In the QGSM calculations we 
accounted for that 
\cite{KhL}
\begin{equation}
\frac{d\sigma}{d\eta^*} = \frac{d\sigma}{dy^*} \cdot \frac{cosh \, \eta^*} 
{\sqrt{m_T^2/p_T^2 + sinh^2 \, \eta^*}}\, ,
\end{equation} 
where the pseudorapidity variable $\eta^*$ can be expressed in terms of the c.m. 
rapidity variable $y^*$ by
\begin{equation}
\eta^* = \frac12 \ln \left[\frac{\sqrt{m_T^2 cosh^2 \, y^* - m^2} + 
m_T sinh \\\, y^*} 
{\sqrt{m_T^2cosh^2 \, y^* - m^2} - m_T  sinh \\\, y^*}\right] \, .
\end{equation} 
The dotted curves in Fig.~7 correspond to the upper limit of 90\% confidence 
level errors. The agreement of QGSM results with experimental points is better 
than 10\%. 

\begin{figure}[h]
\centering
\includegraphics[width=.5\hsize]{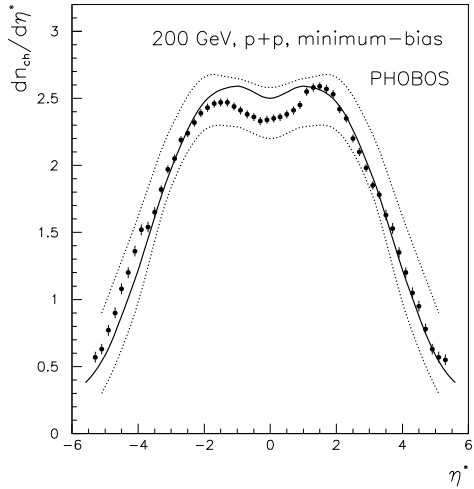}
\vspace*{-0.5cm}
\caption{\footnotesize
C.m. pseudorapidity distribution of charged secondaries produced in minimum 
bias p+p collisions at $\sqrt{s} = 200$ GeV \cite{PHOB}. The QGSM 
calculation is shown by solid curve.}
\end{figure}

The RHIC experimental data for Au+Au collisions \cite{Phob,Phen} give clear 
evidence of the inclusive density saturation effects which reduce 
the inclusive density about two times in the central (midrapidity) region 
in comparison with predictions based on the superposition picture
\cite{CMT,Sh6,APS}. This reduction can be explained in the framework of 
the  inelastic screening corrections  \cite{CKTr}. The effect is very small 
for integrated cross sections (many of them are determined only by geometry), 
but it is very important \cite{CKTr} for the calculations of secondary
multiplicities and inclusive densities at high energies.

Following the estimations in \cite{CKTr}, the RHIC energies are just of
the order of magnitude needed to observe this effect. The inelastic 
screening can make \cite{CKTr} the inclusive density in the midrapidity 
region decrease about two times at RHIC energies and about three times at 
LHC energies in comparison with the calculation without inelastic screening.

However, all estimations are model dependent. The numerical contribution 
of multipomeron diagrams is rather unclear due to the very many unknown 
vertices in the multipomeron diagrams. The number of parameters
can be reduced in some models, for example in \cite{CKTr} the
Schwimmer model \cite{Schw} was used for the numerical estimations.

Another (model dependent) possibility to estimate the contribution of the 
diagrams with Pomeron interaction comes \cite{JUR,BP,JDDSh,BP1} from 
percolation theory. In this approach one assumes that if two or several 
Pomerons are overlapping in transverse space, they fuse in only one  Pomeron. 
When all quark-gluon strings (cut Pomerons) are overlapping, the inclusive 
density saturates, reaching its maximal value at a given impact parameter. 
This approach has only one free parameter $\eta$ called percolation parameter
\begin{equation}
\eta = N_s \frac{r^2_s}{R^2} \langle r(y) \rangle \;,
\end{equation}
with $N_s$ the number of produced strings, $r_s$ the string transverse 
radius, and $R$ the radius of the overlapping area. The factor 
-$\langle r(y) \rangle$ accounts for the fact that the parton density near 
the ends of the strings is smaller that in the central region, where we 
define $r(0) = 1$. At large rapidities we have $N_s$ strings with different 
parton densities, $r_i(y)$, and 
\begin{equation}
N_s \langle r(y) \rangle = \sum_{i=1}^{N_s} r_i(y) \;.
\end{equation}
As a result, the bare inclusive density $dn/dy \vert_{bare}$ is reduced
and we obtain
\begin{equation}
 dn/dy = F(\eta) \cdot dn/dy \vert_{bare} \; ,
\end{equation}
with \cite{BP1}
\begin{equation}
 F(\eta) = \sqrt{\frac{1 - e^{-\eta}}{\eta}} \; .
\end{equation}

For the d+A interaction, which we consider as the sum of p+A and n+A 
interactions (see Eq.~(2)) we obtain at RHIC energy $\sqrt{s_{NN}} = 200$ GeV
for minimum bias interactions $\langle N_d \rangle =$ 1.61, which is
practically the same value as at CERN SPS energy. This value leads to a 
number of participant nucleons, $\langle N_{part} \rangle =$ 8.2, that is 
in agreement with the estimation $8.1 \pm 0.7$ \cite{PHOBOS}. The value of 
$R^2$ in Eq.~(15) is the squared average radius of interactions (nucleon 
radius at not very high energies). The phenomenological estimation of 
$r^2_s$ gives $r_s \sim$ 0.2-0.3 fm  \cite{ABFP} that is in agreement with 
the estimation of the radius of constituent (dressed) quark \cite{AKNS}.

The pseudorapidity spectra of all charged secondaries $dn_{ch}/d\eta^*$
in d+Au collisions at $\sqrt{s_{NN}} = 200$ GeV were measured by PHOBOS 
\cite{PHOBOS} and BRAHMS \cite{BRAHMS} Collaborations. These data allow 
us to obtain the experimental values of $F(\eta^*)$ as (see  Eq.~(17))
\begin{equation}
F_{exp}(\eta^*) = \frac{dn/d\eta^* \vert_{exp}}{dn/d\eta^* \vert_{bare}} \; .
\end{equation}
The corresponding results are presented in Fig.~8 for minimum bias 
\cite{PHOBOS} (left panel) and for 0-30\% central \cite{BRAHMS} (right 
panel) d+Au collisions. Solid lines show the average values of 
$F_{exp}(\eta^*)$ in the interval $-2 < \eta^* < 1$, where the values of 
$F_{exp}(\eta^*)$ have minimal values and are compatible with constant 
behaviour.

\begin{figure}[h]
\centering
\includegraphics[width=.48\hsize]{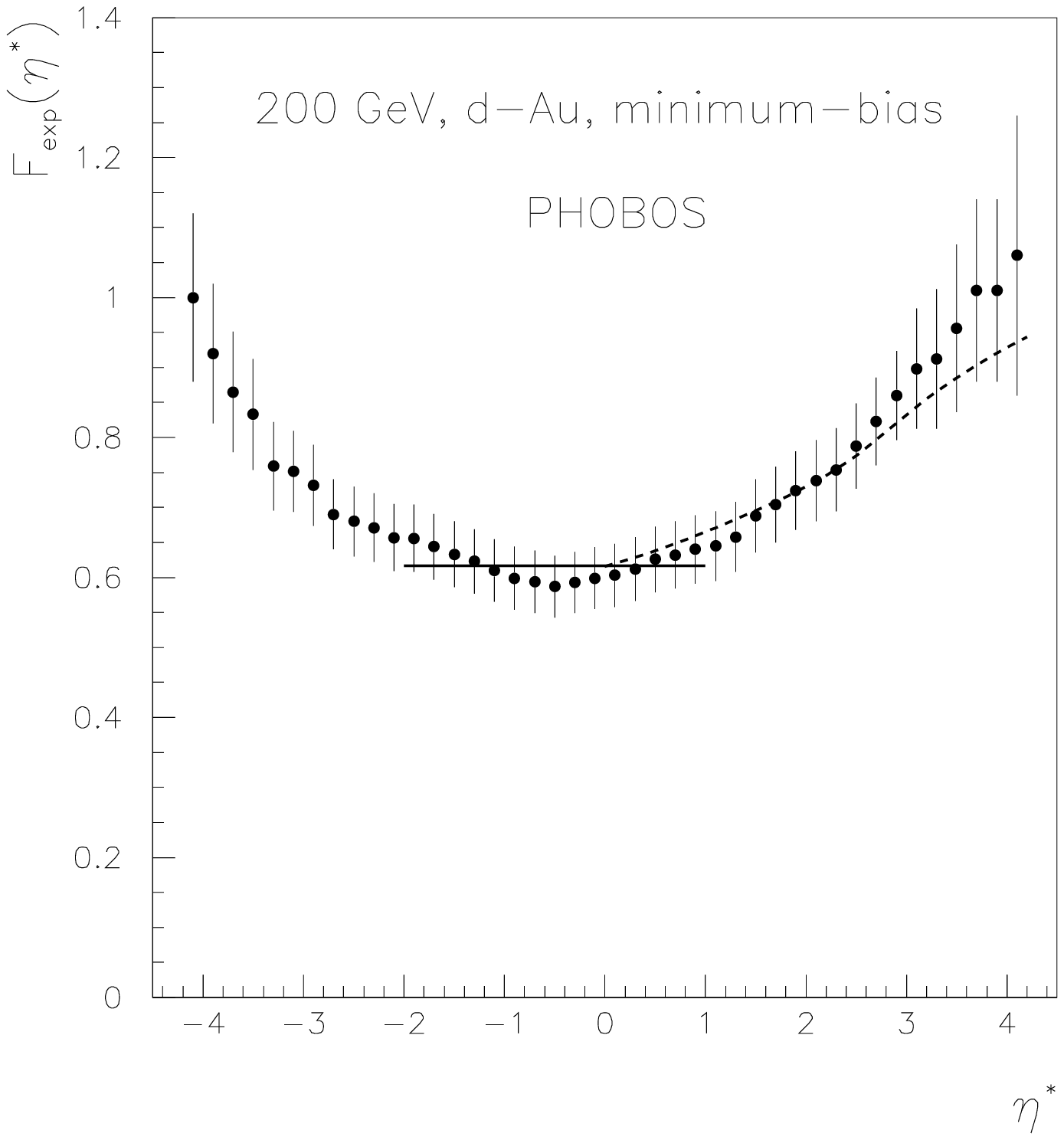}
\includegraphics[width=.48\hsize]{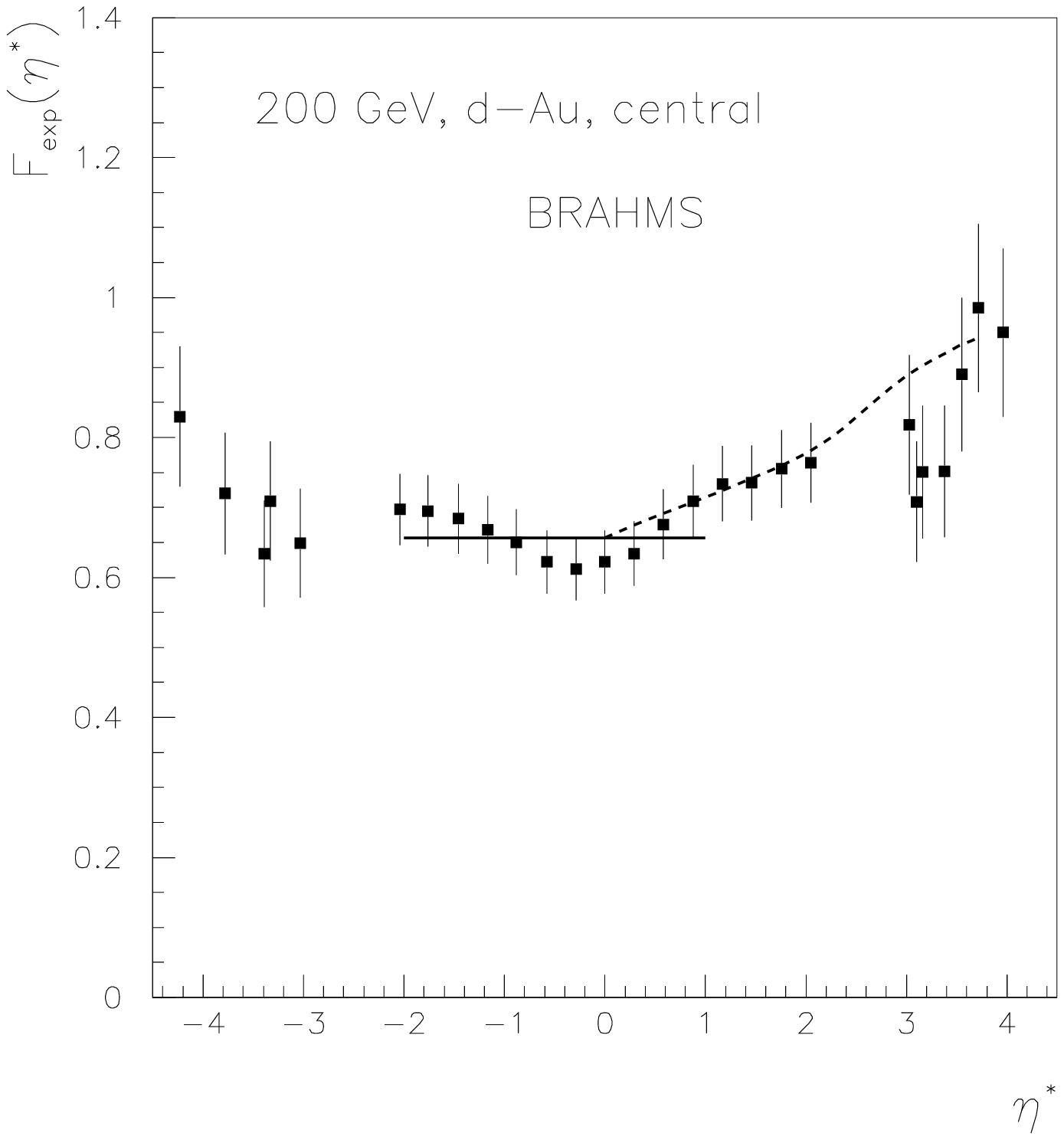}
\vspace*{-0.5cm}
\caption{\footnotesize
C.m. pseudorapidity distributions of the ratios of $F_{exp}(\eta^*)$ for 
charged secondaries at $\sqrt{s_{NN}} = 200$ GeV in minimum bias \cite{PHOBOS}
(left panel) and in 0-30\% central \cite{BRAHMS} (right panel) d+Au 
collisions. Solid lines show the average values of $F_{exp}(\eta^*)$ in the
interval $-2 < \eta < 1$. The dashed curves are the values of 
$F_{exp}(\eta^*)$ using its normalization at $\eta^* =0$.}
\end{figure}

The average value of $F_{exp}(\eta^*)$ in the midrapidity region extracted
from minimum bias events is $F_{exp}(\eta^*) = 0.615 \pm 0.085$, which 
corresponds to the pseudorapidity value $\eta^* = 2.4^{+1.05}_{-0.75}$. These 
error bars are mainly determined by the uncertainty in the calculation of 
$dn/d\eta^* \vert_{bare}$. Now we can estimate the values of
$\langle r(y) \rangle$ (we assume that $\langle r(y) \rangle$ = 
$\langle r(\eta^*) \rangle$ in Eq.~(15)) as
\begin{equation}
\langle r(y) \rangle = \frac{dn(y)/dy \vert_{bare}}
{dn(y=0)/dy \vert_{bare}} \; ,
\end{equation}
and we can calculate the values of $F(\eta^*)$ as a function of pseudorapidity
$\eta^*$  using its normalization at $\eta^* \sim 0$. The results are shown 
in Fig.~8 by a dashed curve and they are in agreement with the experimental 
data.

From the analysis of 0-30\% central d+Au collisions \cite{BRAHMS} (right 
panel in Fig.~8) we obtain the average value of
$F_{exp}(\eta^*) = 0.658 \pm 0.092$, which corresponds to the $\eta^*$ value, 
$\eta^* = 2^{+0.95}_{-0.7}$. In the same way as in the minimum bias case the 
estimation of $\langle r(y) \rangle$ allows us now to calculate the 
pseudorapidity dependence of $F(\eta^*)$ shown in the right panel of Fig.~8 
by a dashed curve.

Finally, from the results shown in Fig.~8 we obtain that the value of the 
percolation parameter $\eta$ for d+Au collisions in the midrapidity region is 
rather large, $\eta \geq 1.5$. This implies a significant contribution 
from the processes with high density parton matter effects. These results
are in qualitative agreement with \cite{KLN}.

Let us now consider the absolute values of pseudorapidity distributions
for produced charged secondaries $dn_{ch}/d\eta^*$. They are presented in 
Fig.~9 for the minimum bias \cite{PHOBOS} (left panel) and 0-30\% central 
\cite{BRAHMS} (right panel) d+Au collisions at $\sqrt{s_{NN}}$ = 200~GeV. 
The standard QGSM calculations without any percolation contributions 
(those in denominator of Eq.~(19)) are shown by solid curves and they are 
in evident disagreement with the experimental data.

\begin{figure}[h]
\centering
\includegraphics[width=.48\hsize]{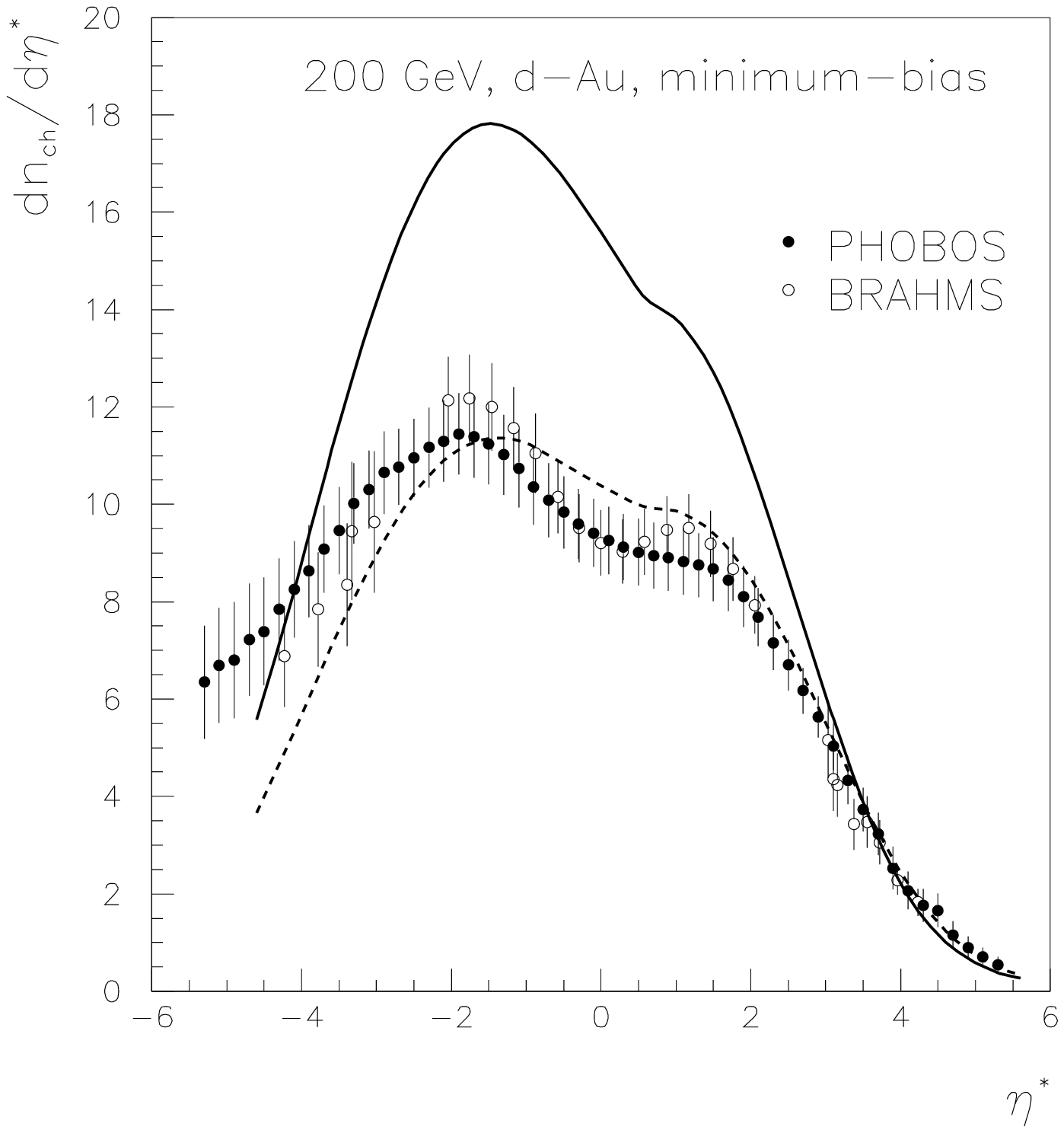}
\includegraphics[width=.48\hsize]{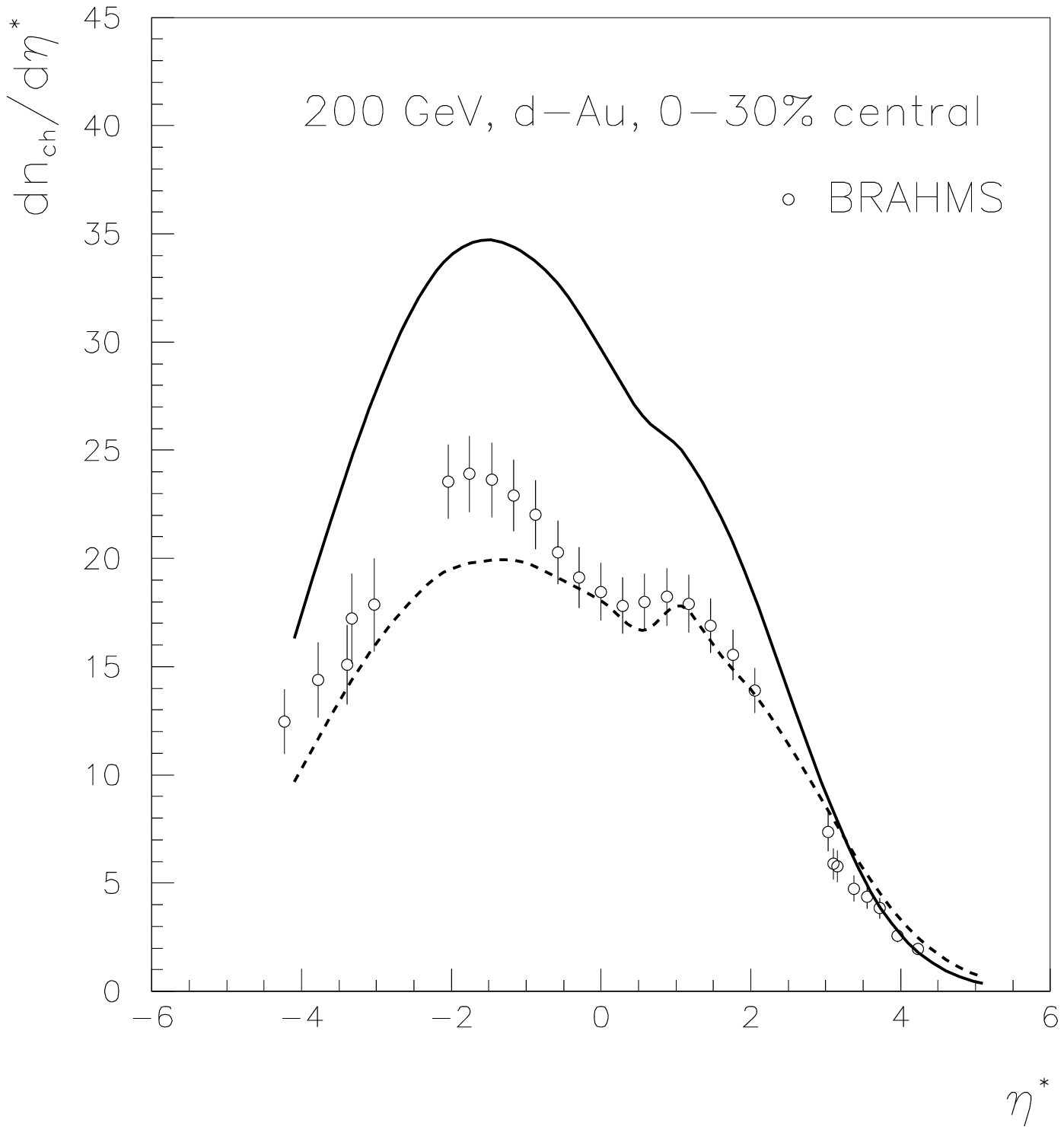}
\vspace*{-0.5cm}
\caption{\footnotesize
C.m. pseudorapidity distributions of charged secondaries at $\sqrt{s_{NN}} = 
200$ GeV measured in minimum bias \cite{PHOBOS,BRAHMS} (left panel) and in 
0-30\% central \cite{BRAHMS} (right panel) d+Au collisions, together with 
their description by the bare QGSM (solid curves), and by taking into account 
percolation contributions (dashed curves).}
\end{figure}

To account for the percolation effects it is technically more simple to 
consider the maximal number of pomerons, $n_{max}$, which can be emitted by 
one nucleon of the deuteron. In this case all model calculations become rather
simple because above the critical value every additional pomeron cannot
contribute to the inclusive spectrum. In this scenario we obtain a 
reasonable agreement with the experimental data (see dashed curves in Fig.~9 
for $n_{max} = 13$). This value of $n_{max}$ is slightly larger than the
average value of produced strings $\langle N_s \rangle$ (see Eq.~(15)).
The last value can be estimated as the double (every cut Pomeron corresponds
to two strings) product of the average value of cut Pomerons in one 
non-diffractive NN interaction $\langle n_{NN} \rangle \sim 2.1$ times the 
average number of interacting nucleons in the Au nucleus 
$\langle \nu_{NA} \rangle = A \sigma_{NN}^{inel} / \sigma_{NA}^{prod} \sim 5$ 
at RHIC energy 
\begin{equation}
\langle N_s \rangle = 2 \langle n_{NN} \rangle \cdot \langle \nu_{NA} \rangle
\end{equation}
(we have used $\sigma_{NN}^{inel} = 43.4$ mb.) So the percolation effects 
do not affect the average configurations of Pomerons, but restrict the 
tails of the distributions.

The yields of $\Lambda$ and $\bar{\Lambda}$ were measured by STAR
Collaboration \cite{STAR} at RHIC energies in both the minimum bias and 0-20\% 
central d+Au collisions at $\sqrt{s_{NN}} = 200$ GeV. In Fig.~10 we compare 
the results of our calculations of $\Lambda$ and $\bar{\Lambda}$
production in minimum bias d+Au collisions with experimental data
\cite{STAR}. The multiplicities of $\Lambda$ and $\bar{\Lambda}$ are 
reproduced in a quite reasonable way. The similar calculations of $\Lambda$ 
and $\bar{\Lambda}$ production at RHIC energy were also presented in 
\cite{BRER}.

\begin{figure}[h]
\centering
\includegraphics[width=.48\hsize]{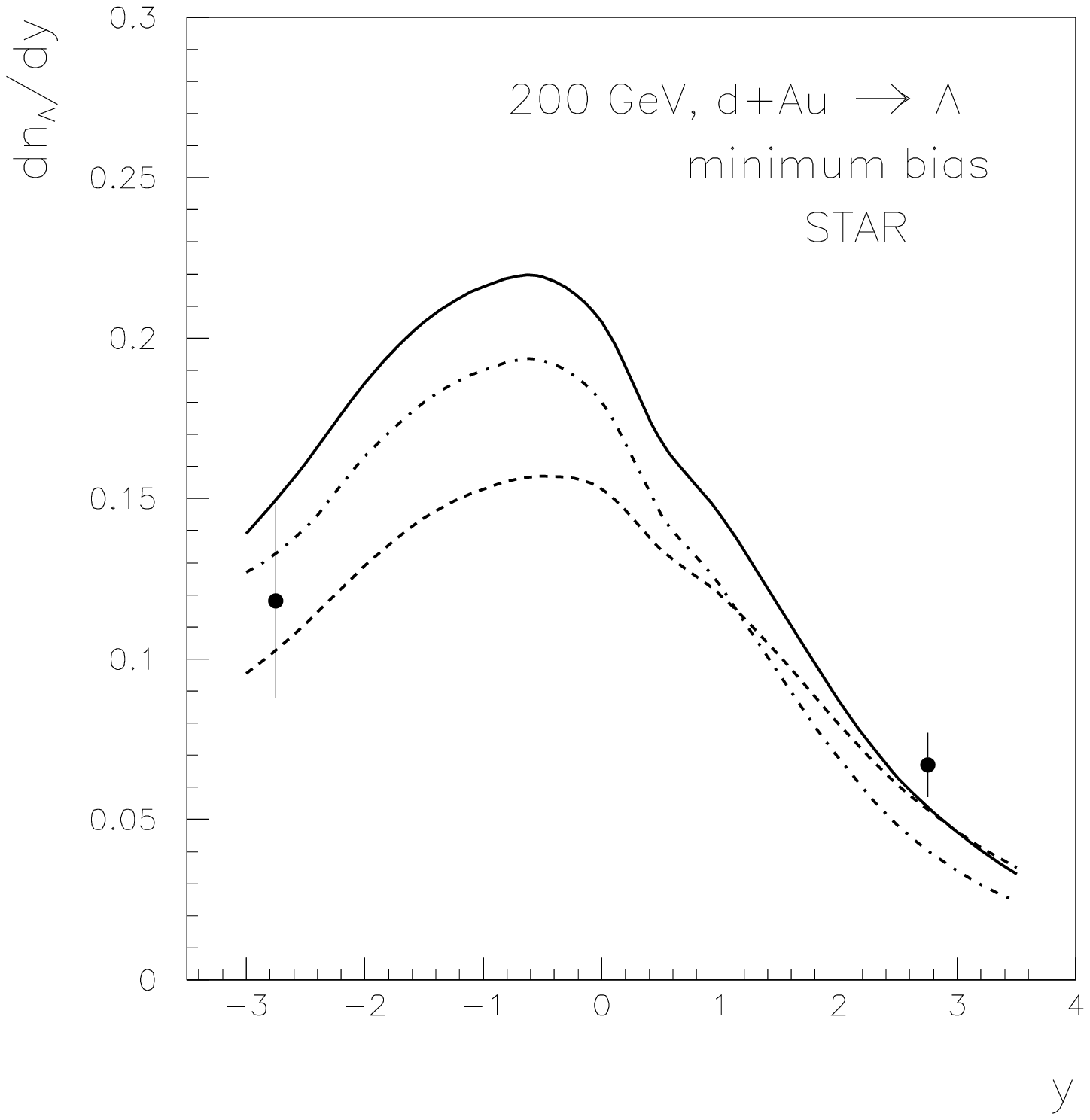}
\includegraphics[width=.48\hsize]{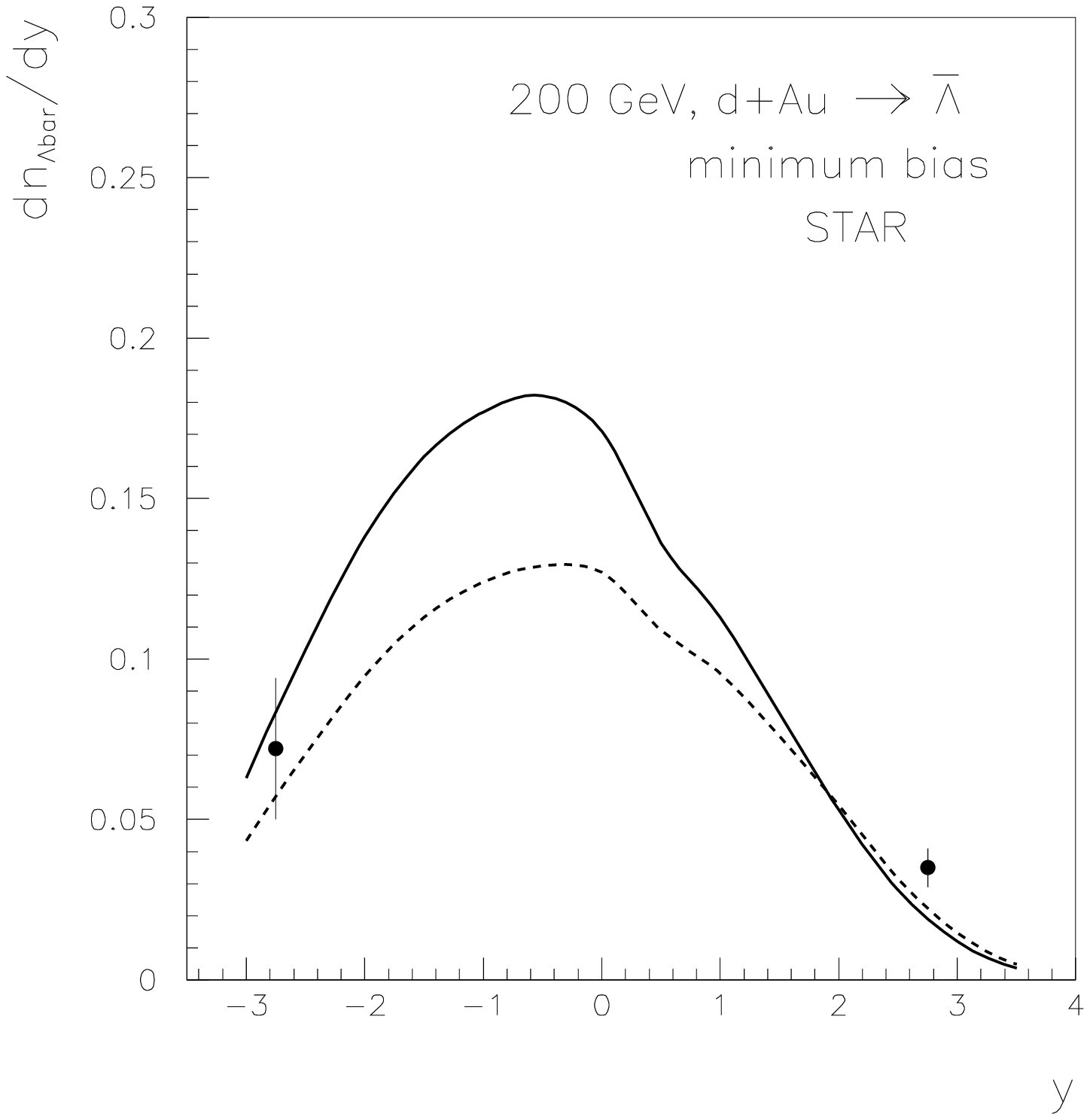}
\vspace*{-0.5cm}
\caption{\footnotesize
Rapidity distributions of $\Lambda$ (left panel) and 
$\bar{\Lambda}$ hyperons (right panel), measured in minimum bias d+Au 
collisions at $\sqrt{s_{NN}} = 200$ GeV \cite{STAR}. The QGSM calculations 
with and without SJ contributions are shown by solid curves and dash-dotted 
curves, respectively. Dashed curves show the results accounting for
percolation effects.}
\end{figure}

The same calculations for $\Lambda$ and $\bar{\Lambda}$ production in
0-20\% central d+Au collisions at $\sqrt{s_{NN}} = 200$ GeV \cite{STAR}
are presented in Fig.~11 where the agreement with the data is also good.

\begin{figure}[h]
\centering
\includegraphics[width=.48\hsize]{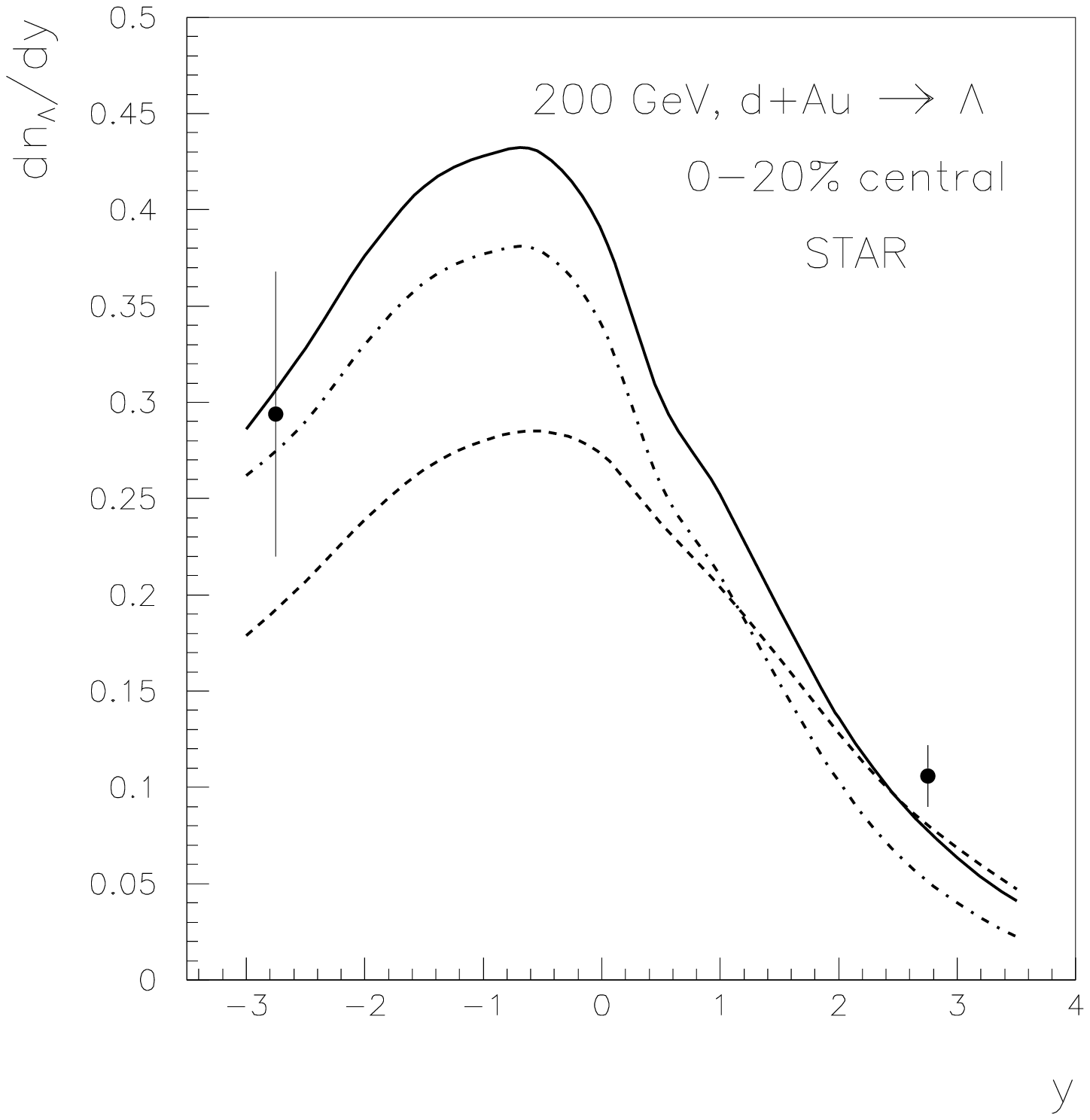}
\includegraphics[width=.48\hsize]{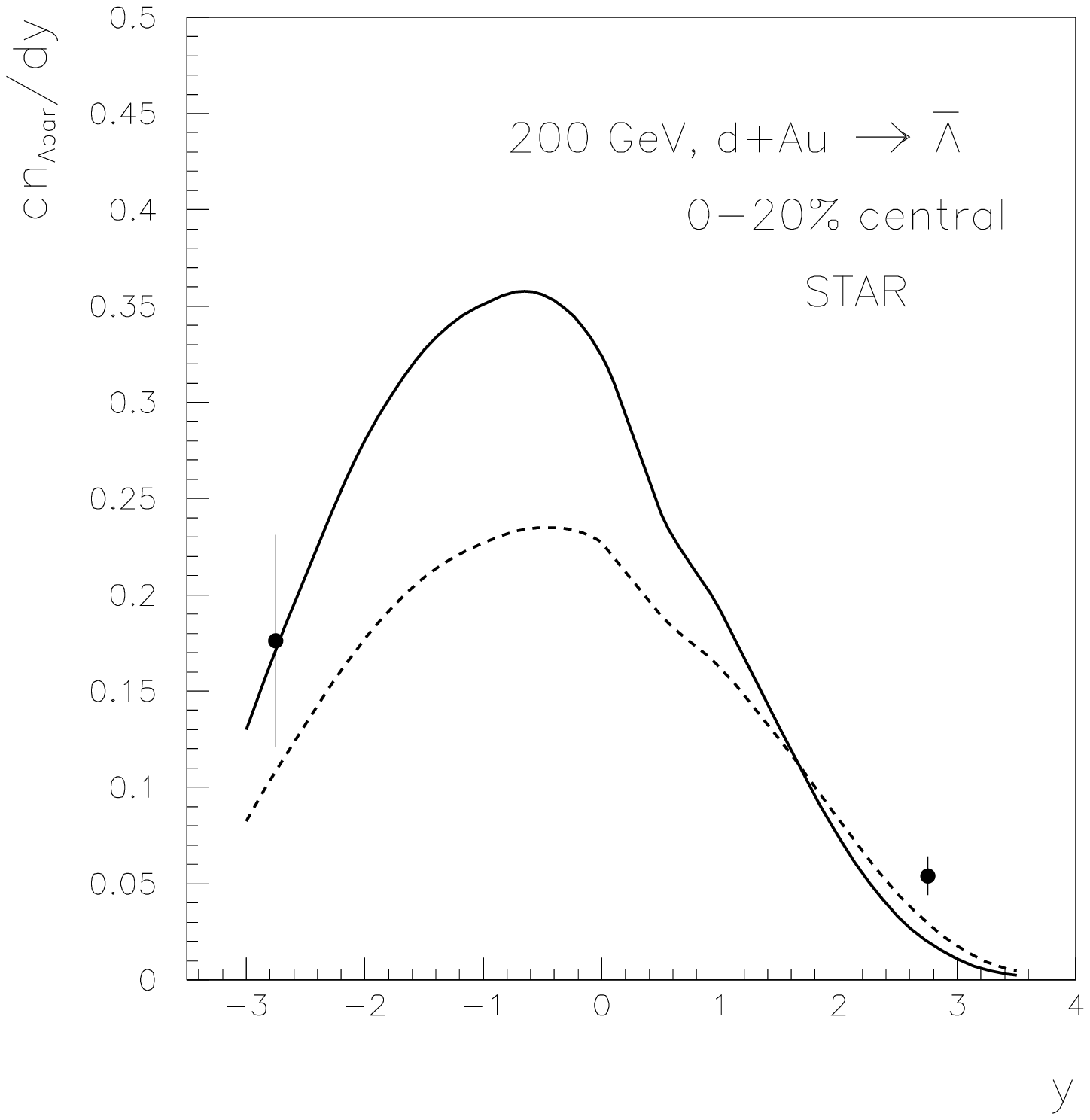}
\vspace*{-0.5cm}
\caption{\footnotesize
Rapidity distributions of $\Lambda$ (left panel) and 
$\bar{\Lambda}$ hyperons (right panel) measured in 0-20\% central d+Au 
collisions at $\sqrt{s_{NN}} = 200$ GeV \cite{STAR}. All curve specifications 
are the same as in Fig.~10.}
-\end{figure}

\section{\bf Predictions for LHC energies}

Let us consider the predictions for p+Au (p+Pb) collisions at LHC energy
$\sqrt{s} = 8.8$ TeV per nucleon.

The calculated result for the charged particle inclusive density in the 
midrapidity region, $dn_{ch}/dy (y=0)$, as a function of initial energy 
without percolation effects is shown in Fig.~12 (left panel) by solid curve. 

\begin{figure}[h]
\centering
\includegraphics[width=.48\hsize]{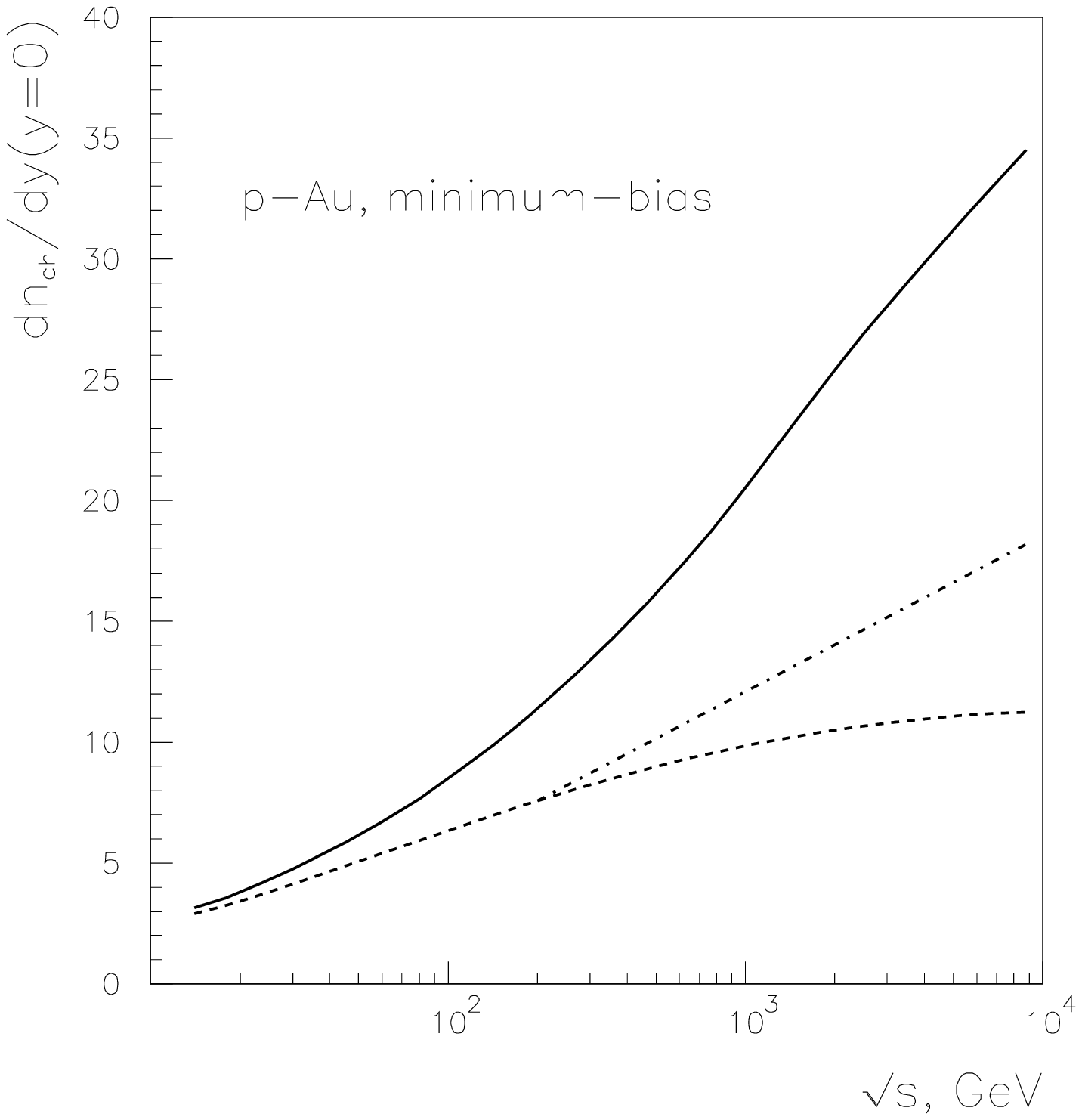}
\includegraphics[width=.48\hsize]{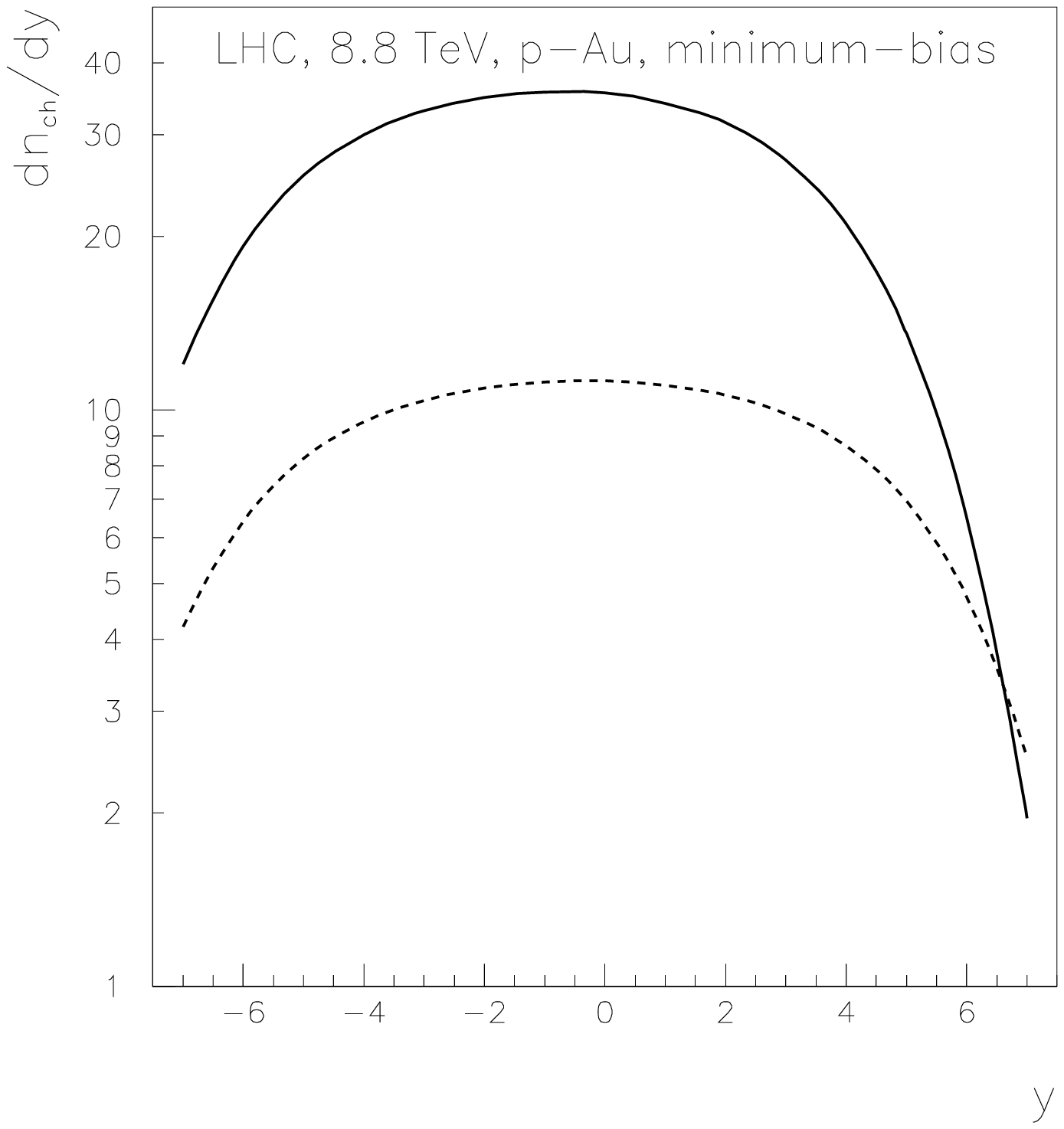}
\vspace*{-0.5cm}
\caption{\footnotesize
The calculated energy dependence of $dn_{ch}/dy (y=0)$ (left panel) and 
the predicted rapidity distributions of charged secondaries in p+Au (p+Pb)
collisions at LHC energy $\sqrt{s_{NN}} = 8.8$ TeV. The results without and
with percolation effects are shown by solid and dashed curves, respectively.
Dash-dotted curve in the left panel shows the result with percolation and with
accounting for an increase of $R^2$ with energy proportional to $\sigma^{inel}_{NN}$ and $r^2_s$ constant
(see Eq.~(15)).}
\end{figure}

In accounting for the percolation effects we meet the problem of the possible 
energy dependence of the maximal number of pomerons, $n_{max}$, which can be 
emitted by the incident proton.

The result obtained in the simplest variant, in which we neglect any energy 
dependence of $n_{max}$, i.e4. we use $n_{max} = 13$ at all energies, is 
shown in the left panel of Fig.~12 by the dashed curve. 

At fixed target energies $\sqrt{s_{NN}} =$ 15-40~GeV percolation effects 
are too small to be observed by comparison of the model calculations with 
experimental data. An additional suppression of the percolation effects at 
fixed target energies comes from the fact that the strings from the nucleus 
need some distance from the nucleus position (lab. frame) in rapidity space 
to fuse (percolation) \cite{Kan} in the midrapidity (c.m. frame) region, but
at relatively low energies this distance is small and percolation effects 
are mainly absent.  

At RHIC energies the percolation effects in $dn_{ch}/dy (y=0)$ are about 
1.5-1.7 times larger then at fixed target energies and they can increase until 
3 times larger at LHC energy. A so large effect in the LHC case is connected 
with the fact that at $\sqrt{s_{NN}} = 8.8$ TeV we have 
$\langle n_{NN} \rangle \sim 3$ and $\langle \nu_{NA} \rangle \sim 8$, so the 
average value of Pomerons ($\sim 24$) in the minimum bias p+A collision is 
significantly larger than $n_{max} = 13$.

The percolation effect for $dn_{ch}/dy (y=0)$ at LHC energy would be smaller 
if we assume that the squared interaction radius $R^2$ in Eq.~(15) increases 
proportionally to the total inelastic $pp$ cross section, whereas the
string transverse radius $r_s$ is constant. In this case, and assuming 
$\sigma^{inel}_{NN}(\sqrt{s_{NN}}$ =8.8 TeV) = 75.5 mb, the value of 
$n_{max}$ can approximately increase proportionally to the
$\sigma^{inel}_{NN}(\sqrt{s_{NN}}$ =8.8 TeV)/$\sigma^{inel}_{NN}
(\sqrt{s_{NN}}$ =200 GeV) ratio and we would 
predict the behaviour shown by the dash-dotted curve in left panel of  Fig.~12.

The predicted rapidity distributions of charged secondaries without (solid 
curve) and with percolation effects with $n_{max} = 13$ (dashed curve) are
presented in the right panel of  Fig.~12. Our solid curve is rather close to 
the estimation of \cite{JDDM} but the shape in the maximum is different.

The processes of baryon number transfer via string junction diffusion 
[24, 28-33] should also be accounted at these so high energies.

%\begin{center}
%{\bf Table}
%\end{center}
%\vspace{15pt}

%The values of $dn/dy/(0.5N_{part})$ for different secondaries produced
%in 5 \% of the most central $Au-Au$ collisions at $\sqrt{s}$ = 130 GeV
%and 200 GeV.

%\begin{center}
%\vskip 12pt
%\begin{tabular}{|c|c|c|c|c|} \hline
%Hadron & \multicolumn{2}{c|}{130 GeV} &
%\multicolumn{2}{c|}{200 GeV}  \\ \cline{2-5}
%& \cite{PHEN1} & QGSM & \cite{PHEN2} & QGSM \\   \hline
%$\pi^+$  & $1.59 \pm 0.05$ & 1.82 & $1.63 \pm 0.13$ & 2.00  \\
%$\pi^-$  & $1.55 \pm 0.05$ & 1.88 & $1.61 \pm 0.13$ & 2.05  \\
%$K^+$   & $0.27 \pm 0.02$ & 0.17 & $0.28 \pm 0.04$ & 0.19 \\
%$K^-$   & $0.23 \pm 0.02$ & 0.17 & $0.26 \pm 0.04$ & 0.18 \\
%p      & $0.16 \pm 0.01$ & 0.13 & $0.10 \pm 0.01$ & 0.14 \\
%$\bar{p}$ & $0.11 \pm 0.01$ & 0.08 & $0.08 \pm 0.01$ & 0.10 \\
%\hline
%\end{tabular}
%\end{center}

%\vskip 0.5cm

\section{Conclusions}

We show that the QGSM together with the Multiple Scattering Theory
can describe on reasonable level the inclusive spectra of secondaries 
produced in d+Au collisions at CERN SPS energies.

The data of RHIC and their comparison with CERN SPS data show numerically 
large effects coming from the inelastic screening effects, or Pomeron 
(secondary particle) density saturation. These effects can be quantitatively 
described in the percolation approach. It is necessary to say that the
scheme used here (the restriction of the number of Pomerons which can be 
cut in one N+A interaction) is different from the approaches used in 
Refs. \cite{JUR,BP,BP1}, but our scheme seems closer to the point of view of 
the Parton Model \cite{Kan,NNN}. The numerical difference with \cite{Kan,NNN} 
comes from the fact that the ratio of $r_s^2/R^2$ in Eq.~(15) is rather small, 
so the percolation parameter, $\eta$, is also small and most of Pomerons can 
exist without percolating. That is why the effects of Pomeron (secondary 
particle) density saturation are small at fixed target energies and they 
become visible only starting from RHIC energies (see left panel of Fig.~12). 
These effects can increase with energy if the ratio of $r_s^2/R^2$ is 
constant, or they can be practically energy independent if the ratio 
of $r_s^2/R^2 \sim 1/\sigma^{inel}_{NN}$.

{\bf Acknowledgements}

We are grateful to N. Armesto and J. Dias de Deus for useful discussions. 
This paper was supported by  Ministerio Educaci\'on y Ciencia of Spain under 
project FPA 2005--01963 and by Xunta de Galicia and, in part, by grants 
RFBR-07-02-00023 and RSGSS-1124.2003.2.

\end {document}